\documentclass[prl,twocolumn,showpacs,preprintnumbers,amsmath,amssymb]{revtex4}

\usepackage{graphicx}
\usepackage{dcolumn}
\usepackage{bm}

\begin{document}

\preprint{}

\title{Manipulation of the quantum state by Majorana transition in spinor\\ Bose-Einstein condensates}

\author{Lin  Xia, Xu Xu, Fan Yang, Wei Xiong, Juntao Li, Qianli Ma, Xiaoji Zhou}
\author{Hong Guo}
\email{hongguo@pku.edu.cn}
\author{Xuzong Chen}%
\email{xuzongchen@pku.edu.cn} \affiliation{School of Electronics
Engineering and Computer Science, Peking University, Beijing 100871,
Peoples Republic of China}%

%

\date{\today}

\begin{abstract}
Manipulation of the quantum state by the Majorana transition in
spinor BEC system has been realized by altering the rotation
frequency of the magnetic fleld's direction. This kind of
manipulation method has no limitation on the transition speed in
principle and the system is well closed, which provides a new and
superior tool to manipulate quantum states. Using this methord on
pulsed atom laser, multicomponent spinor atom laser is generated. We
demonstrate that the experiment results are agreed with the
theoretical predication.
\end{abstract}

\maketitle


Precise and sophisticated control of quantum state is a major goal
in atom physics and is expected to lead to new physics and new
applications, such as quantum computation. An important requirements
of the manipulation process in quantum computation is to be fast (in
order to minimize the effects of decoherence unavoided in real
quantum system, and to speed up the computation). In principle, the
Majorana transition, one kind of nonadiabatic transition, has no
limitation on the transition speed, which has the potential to give
ultra fast manipulation of quantum states. Another advantage of this
manipulation is that the system is well closed if we choose the
ground state of hyperfine structure. This avoids the decoherence
from excited states, such as atoms loses and spontaneous emission.

Majorana transition was first studied in 1932 by Majorana
  \cite{ref12}, who derived the expression of the transition in two level system:
$P_{1/2,-1/2}=\exp(-\frac{f_{Lar}}{f_{Rot}})$
  where $f_{Lar}$ \\[1pt]is the
 Larmor rotation frequency, $f_{Rot}$ the rotation frequency of the
 magnetic field (RFM). After that few quantitative
results on experiments \cite{refb1,refb2} were reported due to
 the limitation of experimental techniques, e.g., the huge velocity distribution of atom beams and the
 nonideal distribution of magnetic field in space.
 The recent experimental realization of Bose-Einstein condensation provides the possibility
 to precisely study Majorana transition since the ensemble can be seen as motionless in the trap.
 Some experiments did show the effects of Majorana
 transition, which are qualitatively observed \cite{ref13,ref14}.
 The observation of population oscillation
 induced by the variation of $f_{Lar}$ was also reproted in
 \cite{ref11}, although the population subjects to the
 position that atoms move to in the magnetic trap, which is not in a real sense
 of manipulation for practical applications.

 In spinor Bose-Einstein condensate system, spin comprises a degree
 of freedom, where the order parameter is a vector rather than a
 scalar in common BEC system. This system offers a new kind of
 coherent matter with complex internal quantum structure and rich
 dynamics, which have been shown in the latest investigation \cite{ref1,ref2,ref3,ref4,ref5,ref6}. As many proposals
 on quantum computation and entanglement based on spinor condensates
 are given \cite{ref7,ref8,ref9,ref10}, experimental manipulation on the spinor freedom degree
 and investigation on dynamics of spinor condensate system
 become particularly important.

In this paper, we introduce how to utilize the Majorana transition
to control quantum states by adjusting $f_{Rot}$, which provides a
powerful tool for quantum states manipulation. The Majorana
transition is precisely studied in spinor BEC system without the
disturbance of atoms huge velocity distribution in early atom beam
experiments \cite{refb1,refb2}. An analytical expression has been
developed and matches the experimental data. The quantum state of
atom laser is manipulated and multicomponent spinor atom laser with
symmetrical population is generated by using this method on pulsed
atom laser. We also show that the Bose-Einstein condensate is most
suitable for studying Majorana transition because low optical
density of images and overlapping of different components will occur
as the temperature increases.

We get samples of condensates in a compact low power
quadrupole-Ioffe-configuration (QUIC) trap with trapping frequency
$\omega_r=2\pi\times220$Hz radially and $\omega_z=2\pi\times20$
axially. The trap consists of a pair of quadrupole coils  which are
assumed to be along $x$-direction, and one Ioffe coil  along
$z$-direction. The direction of gravity is assumed to be
$y$-direction. We design 8mm aperture along the axis of Ioffe coile,
which provides a good optical access for experiments. The shape of
coiles is frustum of a cone in order to provide a larger area to
contact with the water cooled framework. The current is 20.7A in
quadrupole coils and 20.5A in Ioffe coil. Typically, a $^{87}Rb$
condensate with $2\times10^5$ atoms at $|F=2$, $m_{F}=2\rangle$
state is formed after evaporative cooling.

 We redesign the
circuit for the turn-off of coils to control the switching off time
of two type coils separatively. The Ioffe coil is in parallel with a
variable resistor, and the turn-off time of Ioffe coil is set by
$L_I/R_I$, where $L_I$ is the inductance of the coil and $R_I$ is
the resistance of the loop. Conveniently, the turn-off time of Ioffe
coil can be tuned continuously by setting the resistance of the
variable resistor. Similarly, the quadrupole coils are in parallel
with another resistor. To control the switching off time of the
magnetic field separately, the eddy current must be strictly
removed. We redesign the magnetic trap with grooves in the framework
and get rid of the influence of eddy current, which can be shown by
comparing the decreasing process detected by a detective coil with
by the Hall effect current sensor. \setlength{\floatsep}{8pt plus
2pt minus 2pt}
 \setlength{\textfloatsep}{10pt plus 2pt minus
2pt} \setlength{\intextsep}{10pt plus 2pt minus 2pt}
\begin{figure}[h]
\setlength{\abovecaptionskip}{-0pt}
\setlength{\belowcaptionskip}{0pt}
\begin{center}
\includegraphics[width=7.5cm]{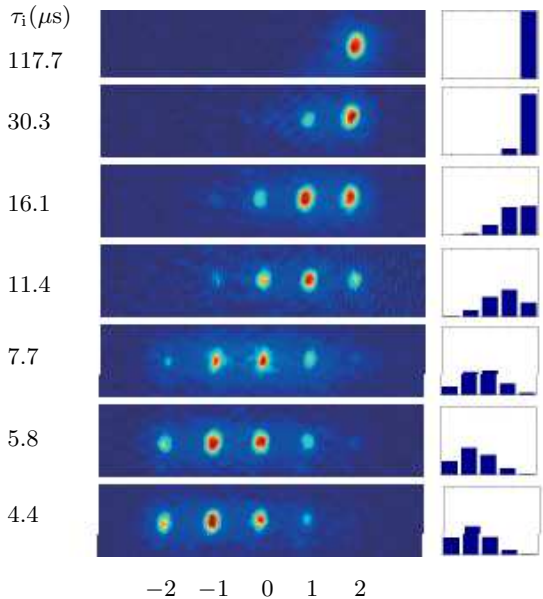}
\put(-207,202){ $\mathrm{\tau_{i} (\mu s)}$}
\put(-205,185){117.7}\put(-205,162){30.3}\put(-205,131){16.1}\put(-205,100){11.4}\put(-205,73){7.7}\put(-205,42){5.8}\put(-205,13){4.4}
\put(-153,-15){$-2$}\put(-133,-15){$-1$}\put(-109,-15){$0$}\put(-92,-15){$1$}\put(-74,-15){$2$}
\end{center}
\caption{\label{label}Observation of different components vs the
turn-off time of Ioffe coil: The atoms are initially prepared in
$|F=2$, $m_F=2\rangle$ state. The five parts of the condensates from
right to left, also the same as the following figures, refer to the
five states $|2,2\rangle$, $|2,1\rangle$, $|2,0\rangle$,
$|2,-1\rangle$ and $|2,-2\rangle$ respectively. The dimension of
each picture is $1.2\rm mm\times 0.3mm$. During the process, the
turn-off time of quadrupole coils is fixed at $\tau_q=117.7\rm \mu
s.$ We also show the population distributions obtained from the
analytical expression in the bar charts.}
\end{figure}

When the nonsynchronous decreasing process of the magnetic field
from different coils emerges, the condensate atoms will experience
the zero field of $B_z=0$ and the reversion of the direction. This
is easy to understand from the formation of the magnetic field. We
know that the Ioffe coil's field is in the opposite direction of
that of quardupole coils along $z$ direction, and the total field's
direction is along the Ioffe's field. If the Ioffe's field decays
fast, the zero field emerges and the field's direction reverses in
$z$ direction. In this case, Majorana transition happens and the
different components of BEC will be separated in space by the
Stern-Gerlach effect induced by the gradient of the magnetic field.
If the turn-off speed of Ioffe coil increases, the reversion speed
of field increases too, which induces the increasing of $f_{Rot}$.
Then more spinor components are transited from initial component
$|F=2$, $m_{F}=2\rangle$. Setting $\tau_i$, the RFM $f_{Rot}$ at the
time of reversion is uniquely determined as $\tau_q$ is a constant,
where $\tau_i$ and $\tau_q$ are the $1/e$ turn-off time of Ioffe
ciol and quadrupole coils, respectively. Fig. 1 shows the relation
between Majorana transition and turn-off time of the Ioffe coil in
experiments. The distribution of different spinor components
subjects to Majorana transition, which is controlled by adjusting
the turn-off time of Ioffe coil. The distribution of spinor
components well fits the theoretical prediction (see  Fig. 1 right
charts), which is described by Eq. (\ref{eq1}) and (\ref{eq13}).
When $\tau_i=\tau_q$, the magnetic field does not reverse in $z$
direction and the single component condensate in $|F=2$,
$m_{F}=2\rangle$ forms (see Fig. 1). As $\tau_i$ decreases, the
$f_{Rot}$ increases and the induced Majorana transition becomes
stronger.

To investigate the Majorana transition process in experiments more
accurately, we give the theoretical analysis in the following.
Majorana studied the transition in the model where the magnetic
field evolves as $B_x(t)=0$, $B_y(t)=\rm const$, $B_z(t)=Kt$ and
$t=(-\infty, \infty)$. The Majorana formula has been derived both
from quantum mechanics and group theory \cite{ref12}\cite{ref17}.
For
multilevel system with a total angular moment $J$,\\[-10pt]
\begin{eqnarray}\label{eq1}
P_{m,m'}=(J+m)!(J+m')!(J-m)!(J-m')!\left(\cos\frac{\theta}{2}\right)^{4J}
                                                            \nonumber\\
\times\left[\sum_{\nu=0}^{2J}\frac{(-1)^\nu(\tan\frac{\theta}{2})^{2\nu-m+m'}}{\nu!(\nu-m+m')!(J-m-\nu)!(J-m'-\nu)!}\right]^2,
\end{eqnarray}\\[-10pt]where the value of $\theta$ is given by the two level transition\\[-10pt]
\begin{displaymath}
\sin^2\left(\frac{\theta}{2}\right)=P_{1/2,-1/2}.
\end{displaymath}\\[-10pt]Since the result to the system with arbitrary angular momentum $J$ can be
generalized from the two level system, we emphasize on the two level
case.

For a motionless system during the transition process with spin
moment $s=1/2$ in magnetic field $B(t)$, the Schr\"odinger equation
can be written as\cite{ref11}\\[-28pt]

\begin{equation}\label{eq2}
i\hbar\left(\begin{array}{ccc}
\dot{c_1}\\
\dot{c_2}\\\end{array}\right)
 =\frac{g\mu_B}{\hbar}\hat{\vec{F}}\cdot\vec{B}(t)\left(\begin{array}{ccc}
c_1 \\
c_2 \\\end{array}\right),
\end{equation}\\[-10pt]where $\hat{\vec{F}}=\hbar/2\hat{\vec{\sigma}}.$ The magnetic field
which atoms experience takes the form $\vec{B}(t)=[0, B_y(t),
B_z(t)]$, where $B_x(t)=0$ is produced by the symmetry of the trap
and $B_y(t)$ is produced due to the dragging down of the gravity.
The experimental evolution of the magnetic field due to the
discharging process of coils is exponential.

 In our case, the variable $\tau_i$ is
much less than the constant $\tau_q$ when Majorana transition
happens obviously. By taking approximation, it is appropriate to
write $\vec{B}(t)$ as\\[-35pt]

\begin{eqnarray}\label{eq11}
B_y(t)&\approx &A_{y},\nonumber\\
B_z(t)&\approx &A_z-C_zt,
\end{eqnarray}\\[-20pt] where\\[-25pt]
\begin{eqnarray}\label{eq7}
A_y&=&B_{yI}+B_{yQ},\nonumber\\
A_z&=&B_{zI}-B_{zQ},\nonumber\\
C_z&=&B_{zI}/\tau_i-B_{zQ}/\tau_q.\nonumber
\end{eqnarray}\\[-15pt]  $B_{yI}$ and $B_{yQ}$ are the initial magnetic field respectively
generated by Ioffe coil and quadrupole coils in $y$ direction,
$B_{zI}$ and $B_{zQ}$ are in $z$ direction. The expressions of the
RFM and Larmor frequency are $f_{Rot}=\partial B(t)/2\pi
B(t)\partial t$ and  $f_{Lar}=g\mu_0 B(t)/2\pi\hbar$, respectively.
In the case of Eq. (\ref{eq11}), $f_{Rot}=C_z/2\pi A_y$ at the time
when the magnetic field reverses its direction. From the expression
of $C_z$, we can see that the rotation frequency of the magnetic
field is only determined by $\tau_i$ as $\tau_q$ and $A_y$ are
fixed. By substituting Eq. (\ref{eq11}) into Eq. (\ref{eq2}), the
second-order differential equations which can be transformed into
Webber equations \cite{ref18} are obtained. The asymptotic solution
of the Webber
equation corresponding to the experiments can be derived as\\[-20pt]
\begin{equation}\label{eq13} P_{1/2,-1/2}
=\exp\left(-K\frac{A_y^2}{\frac{B_{zI}}{\tau_i}-\frac{B_{zQ}}{\tau_q}}\right)
\end{equation}\\[-9pt]
where $K$ is a constant. The population of different spinor quantum
states can be controlled by adjusting the switching off time
$\tau_i$ in Eq. (\ref{eq13}), which is the only adjustable
parameter. This expression reveals the essence of the Majorana
transition observed generally in experiments.

\begin{figure}[h]
\setlength{\abovecaptionskip}{-10pt}
\setlength{\belowcaptionskip}{0pt}
\begin{center}
\includegraphics[width=7.0cm]{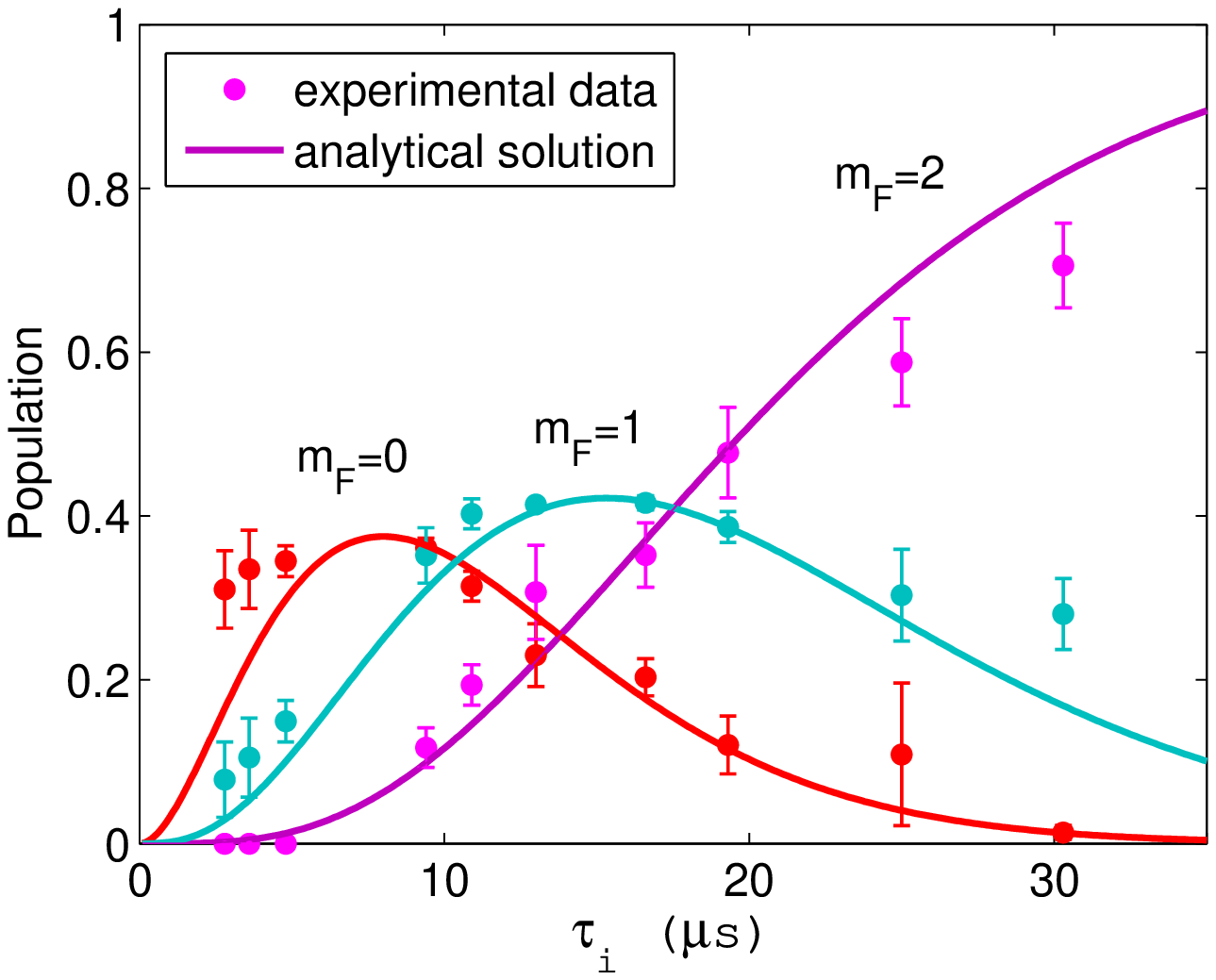}
\includegraphics[width=7.0cm]{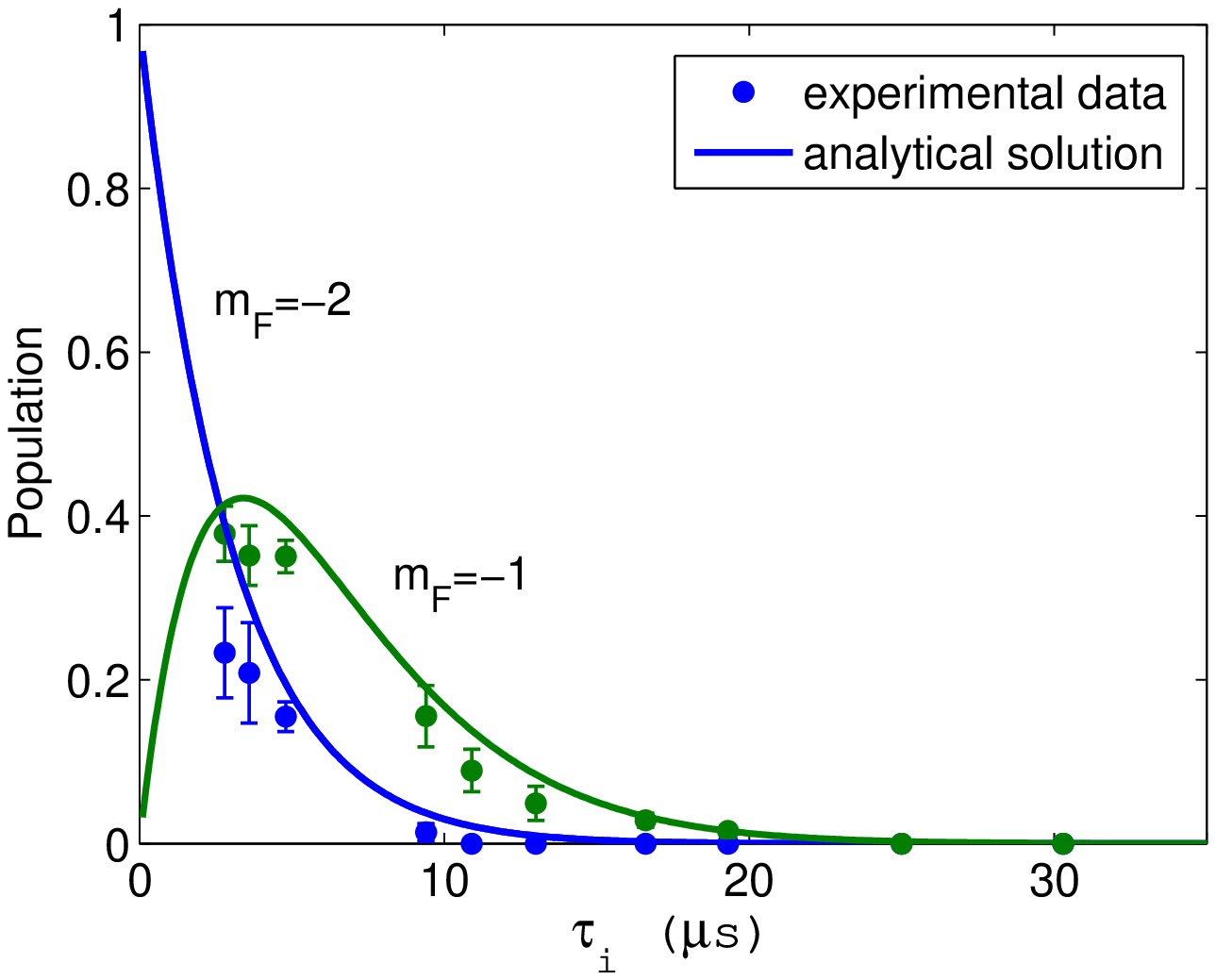}
\end{center}
\caption{\label{label}Evolution of population as the turn-off time
of Ioffe coil $\tau_i$ with atoms initially in $|F=2$, $mF=2\rangle$
state: The experimental data and analytical solution are shown.}
\end{figure}

The comparison between the theoretical prediction and the
experimental results based on the experiments of Fig. 1 is shown in
Fig. 2, which describes the evolution of spinor BEC with different
turn-off time of Ioffe coile. The experimental data and theoretical
results with atoms initially prepared in $|F=2,m_F=2\rangle$ state
are shown and match each other. We can see that the population is
initially prepared in $m_F=2$ and then transited to $m_F=-2$ state
gradually through the $m_F=1, 0$ and $-1$ state. The asymptotic
state as $\tau_i$ decreases is $m_F=-2$ state. This corresponds to
the total nonadiabatic process, in which the magnetic field reverses
instantaneously so that the magnetic moment maintains its direction.
The final distributation of atoms initially prepared in any spinor
state can be easily derived by using Eq. (\ref{eq1}) and
(\ref{eq13}). Similarly to optical pumping and Rabi oscillation, the
Majorana transition is another choice to manipulate the quantum
state transition by controlling the turn-off time in experiments.

\begin{figure}[h]
\setlength{\abovecaptionskip}{-10pt}
\setlength{\belowcaptionskip}{0pt}
\begin{center}
\includegraphics[width=7.0cm]{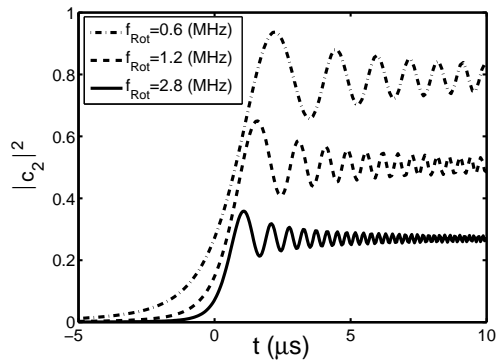}
\end{center}
\caption{\label{label}Evolution of population as time in Majorana
transition: Plotted is the evolution of $\mid c_2\mid ^2$ as time
for the three values of magnetic field rotation frequency
$f_{Rot}=0.6$ MHz, $1.2$ MHz and $2.8$ MHz.}
\end{figure}

Fig. 3 shows the time evolution of the population by numerically
solving the Webber equations. In \cite{ref11}, the transition period
is estimated to be only about $1\mu s$ by comparing the Larmor
frequency with the field rotation frequency. Fig. 3 is another and
direct way to show the transition period, which is consistent with
the estimation in \cite{ref11}. So the transition period length can
be approximately estimated as the period length of $f_{Rot}\ge
f_{Lar}$, whose expression is
\begin{displaymath}\ \triangle
t=2\cdot\sqrt{\left(\frac{B_y\hbar}{g\mu_B}\right)^{2/3}C_z^{-4/3}-B_y^2C_z^{-2}}
.\end{displaymath} In the expression of $\triangle t$, the
transition period is monotone decreasing and the asymptotic value is
$0$ (no limitation on the transition speed) at the region where
$C_z$ tends to infinite. This trend also appears in Fig. 3.

\begin{figure}[h]
\setlength{\abovecaptionskip}{-0pt}
\setlength{\belowcaptionskip}{0pt}
\begin{center}
\includegraphics[width=7.0cm]{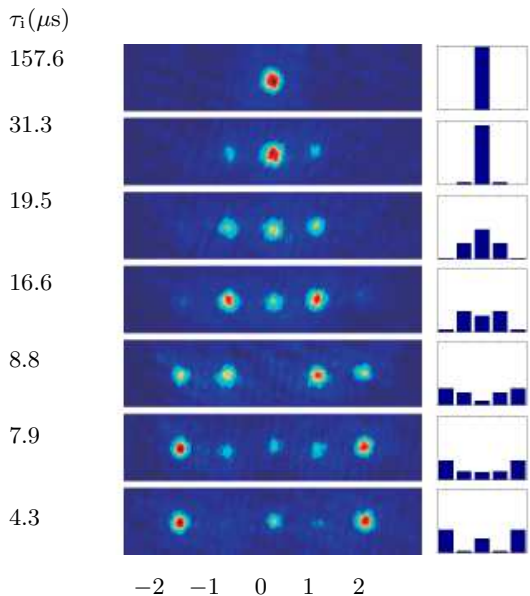}
\put(-203,200){ $\mathrm{\tau_{i}(\mu s)}$}
\put(-200,185){157.6}\put(-200,160){31.3}\put(-200,131){19.5}\put(-200,100){16.6}\put(-200,71){8.8}\put(-200,42){7.9}\put(-200,11){4.3}
\put(-153,-15){$-2$}\put(-132,-15){$-1$}\put(-107,-15){$0$}\put(-89,-15){$1$}\put(-70,-15){$2$}
\end{center}
\caption{\label{label}Generation of multicomponent spinor atom laser
by Majorana transition: The $m_F=0$ spinor state is prepared by
producing pulsed atom laser with rf radiation. The transition
process is identical to that in Fig. 1. The population distributions
obtained from the analytical expression are shown in the bar
charts.}
\end{figure}

The generation of multicomponent spinor atom laser by Majorana
transition is shown in Fig. 4. We obtain atoms prepared in $m_{F}=0$
spin state by producing pulsed atom laser with short rf radiation.
The total number of atoms prepared in $m_{F}=0$ state are about
$4\times 10^4$. After the $m_{F}=0$ atom laser is generated and
falls down for 2 ms, the nonsynchronous process of the magnetic
field happens and the multicomponent spinor atom laser is generated.
We can manipulate the distribution of the population by adjusting
$\tau_i$, which is the same as atoms initially prepared in $m_{F}=2$
state. Fig. 4 shows the symmetry of the population distrubition.
This agrees with Eq. (\ref{eq1}) when atoms are initially prepared
at $m_F=0$ state. The generation of multicomponent spinor atom laser
indicates that Majorana transition is a powerful tool for quantum
state manipulation.

\begin{figure}[h]
\setlength{\abovecaptionskip}{-0pt}
\setlength{\belowcaptionskip}{0pt}
\begin{center}
\includegraphics[width=4.5cm]{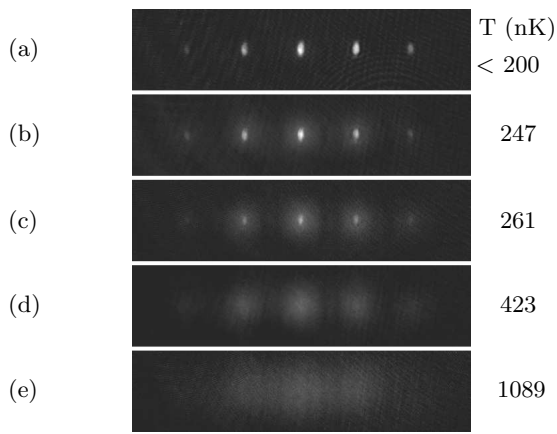}
\put(-175,142){(a)} \put(-175,110){(b)}
\put(-175,77){(c)}\put(-175,45){(d)}\put(-175,13){(e)}\put(3,150){T
(nK)}\put(2,136){$<$ 200}\put(8,110){ 247} \put(8,77){
261}\put(8,45){ 423}\put(7,13){ 1089}
\end{center}
\caption{\label{label}Majorana transition with atoms at different
temperature. From (a) to (e), the maximum optical density is 0.95,
0.91, 0.57, 0.27 and 0.23, respectively.}
\end{figure}

In addition, the Majorana transition  with atoms at different
temperature is studied (see Fig. 5). The temperature of atoms is
altered by setting the end frequency of evaporative cooling. As the
temperature rises, the detected maximum optical density decreases,
e.g., 0.95 in (a) yet 0.23 in (e). In fact, it is hard to get
information of atoms in (e) because of the bad contrast. The lowest
temperature of sub-Doppler cooling in Rb is $4$ $\rm \mu K$
\cite{ref19} which is much higher than the temperature in (e). Even
at about the critical temperature [see Fig. 5(d)], the small
fractions of $m_F=2$ and $m_F=-2$ can hardly be detected. Another
problem for atoms at high temperature is the overlapping of
different component atoms [see Fig. 5(e)]. For these two reasons,
BEC appeals to be the most suitable tool to study Majorana
transition.

In conclusion, this work establishes a basic tool to control the
macroscopic coherent quantum state of spinor condensates for
developing more sophisticated control. As an application,
multicomponent spinor atom laser is generated. The Majorana
transition is precisely studied without the disturbance of atoms
velocity distribution. Our treatment for Majorana transition leads
to a quantitative agreement with the experimental results. With our
model, the dynamical evolution of atoms prepared in any spinor state
in variable magnetic field can be described. We also show the
superiority of Bose-Einstein condensation as a tool to study
Majorana transition.

\section{\label{sec:level}acknowledgments}
The authors thank Professor J\"org Schmiedmayer at Heidelberg
University and Professor Li You at Georgia Institute of Technology
for their helpful discussions. This work was supported by the
National Fundamental Research Programme of China under Grant No.
2001CB309308, the Major Program of National Natural Science
Foundation of  China under Grant No. 60490280, National Natural
Science Foundation of China under Grant No. 60271003 and 10474004.



\end{document}